# Multi-task Prediction of Patient Workload

Mohammad Hessam Olya
Department of of Computer Science
Wayne State University
Detroit, USA
h.olya@wayne.edu

Dongxiao Zhu
Department of Computer Science
Wayne State University
Detroit, USA
dzhu@wayne.edu

Kai Yang
Department of Industrial and Systems Engineering
Wayne State University
Detroit, USA
kai.yang@wayne.edu

*Abstract*— Developing reliable workload predictive models can affect many aspects of clinical decision-making procedure. The primary challenge in healthcare systems is handling the demand uncertainty over the time. This issue becomes more critical for the healthcare facilities that provide service for chronic disease treatment because of the need for continuous treatments over the time. Although some researchers focused on exploring the methods for workload prediction recently, few types of research mainly focused on forecasting a quantitative measure for the workload of healthcare providers. Also, among the mentioned studies most of them just focused on workload prediction within one facility. The drawback of the previous studies is the problem is not investigated for multiple facilities where the quality of provided service, the equipment, and resources used for provided service as well as the diagnosis and treatment procedures may differ even for patients with similar conditions. To tackle the mentioned issue, this paper suggests a framework for patient workload prediction by using patients' data from 130 VA facilities across the US. To capture the information of patients with similar attributes and make the prediction more accurate, a heuristic cluster-based algorithm for single-task learning is developed in this research. Moreover, most of the time demand forecasting is associated with the independence assumption between tasks. To the best of our knowledge, the previous studies in the area of patient demand forecasting merely considered task relatedness. In this research, we have considered patient-dependent and facility-dependent attributes and the relation between tasks into the model while implementing Multi-Task Learning (MTL) approach and training multiple related tasks simultaneously. The results of this study show that incorporating the relatedness of the tasks into the patient workload prediction model, leads to a higher performance compare to ignoring the relationship between multiple related tasks while doing single-task learning.

I. INTRODUCTION

Uncertainty in patient's demand and the workload is one of the critical factors that tremendously affects healthcare delivery systems. The source of patient's demand variation is rooted in natural differences of individuals within a population who are different in terms of socioeconomic factors and health conditions. Demand prediction helps decision makers to estimate the required healthcare resources more accurate in order to manage the risk and workload of healthcare providers by providing them insight into developing healthcare intervention strategies. Specifically, underestimating the demand can affect the quality of the provided care adversely, whereas overestimating demand increases operating costs. Although there are lots of approaches adapted recently in order to forecast the patient demand, there is still room for improvement in healthcare demand prediction due to various types of uncertainty and patterns exist in these types of problems.

Reliable data and proper analytical approach are the essentials of patient demand prediction and developing data-driven models [1]. Electronic Health Records (EHR) are considered as useful sources of structured data that enable investigating the effect of different clinical attributes and relationship between various types of diseases and comorbidities with the patient demand by applying machine learning and statistical analysis tools on clinical data. Data analysis techniques transformed the healthcare research on heterogenous patient information. As it is mentioned, EHR plays a vital role as the primary source for healthcare data analytics. EHR strengthen the research in healthcare by capturing and integrating patient-specific information and providing a high dimensional data that includes diagnosis results (ICD codes), patient condition, treatments, medications, laboratory test results and imaging data as well as billing information, socioeconomic and demographic data such as age, gender, and employment status. EHR is a basis for building data-driven predictive clinical multivariate models using machine-learning methods to make inferences on patient demand based on their different features and attributes.

Nowadays healthcare systems are transforming from disease centered systems to patient-centered and team-based systems. This process necessitates the integration of healthcare providers, patients, and medical facilities. However, there are many challenges involved in quantifying the performance, productivity and measuring the healthcare providers' amount of work as well as defining billing, payment and compensation procedures based on the actual workload of the providers. EHR contains various types of data among which relative value unit is considered as one of the most valuable information since it helps in to quantify the workload that patient imposes on each healthcare provider. Using RVUs instead of number of billings or patients, to assess the amount of workload is considered as an essential component of numerous physician practices recently.

Many federal programs and private payers make use of this unique methodology as a basis for healthcare team compensation calculations in modern healthcare service industry. Physician work RVU is the relative measure of time needed to complete the service, technical skill and physical effort, intensity, the level of mental effort and training as well as judgment required to supply a specified health service.

RVUs are selected based on the Healthcare Common Procedure Coding System (HCPCS) which is an accredited healthcare procedure codes defined by the American Medical Association's Current Procedural Terminology (CPT) for reporting hospital procedures in different levels. Physicians create a report of all the services and their associated codes performed on every patient. These codes represent which course of action was performed, which treatments have been prescribed, injected, or delivered to the patient. The CPT codes ought to be in line with the doctor 's diagnosis which is represented by International Classification of Diseases (ICD) codes. These codes also represent a doctor 's diagnosis as well as the patient 's condition. RVU also represents the relativeness of values assigned to different services. Every single type of service or procedure is given a specific amount of RVU. More complicated procedures are given larger RVU. For instance, a well patient visit would be given a lower RVU than an invasive surgical procedure. Given this relative scale, a medical practitioner visiting five complicated or high acuity patients per day may accumulate far more RVUs compared to a physician who visits ten or more low acuity patients per day. As it is mentioned a code with a higher RVU work requires more time, technical skills and intensity. For instance, a level one office visit could be given an RVU of one, a quality three office visit may be given an RVU of two, and surgical treatment may be assigned an RVU of twenty. Thus, patients' different conditions and diagnosis can affect the workload of the healthcare providers. Therefore, in order to reduce the cost of health systems and enhance patient satisfaction, developing a predictive model to estimate the workload is an essential task.

As it is mentioned in this study, we are dealing with EHR from many healthcare facilities. Multi-task learning is a useful solution to approach datasets that contain multiple related instances. The goal of MTL is to manage helpful information in multiple similar tasks in order to enhance the generalization performance of all of the tasks. To be able to improve the performance of learning tasks, MTL may be integrated with other learning models including semi-supervised learning and unsupervised learning. The large number of the samples is an essential requirement for machine learning approaches in order to learn a precise learner. However due to difficulty in gathering the healthcare data, in healthcare systems analytics this requirement might not be satisfied easily [2]. In the case that there is a limited number of samples for each task and majority of learning tasks are related to each other, learning these tasks mutually can lead to improved training performance in comparison with learning them individually. MTL categorizes the dataset based on specific tasks and considers a limited training dataset for each defined task then learn all of them jointly to use the shared representation; what is learned for each task can improve the performance of learning other tasks. Healthcare demand depends on many facility dependent features that can be different across various facilities. Thus, facilities can be considered as different tasks with their own data in multi-task learning while learning them jointly to develop a precise learner.

In this paper, we concentrate on patient workload predication which is an important and critical issue in any healthcare system by implementing multi-task learning approach on patient's data to achieve an accurate demand prediction. The accuracy and dependability of demand prediction is crucial in healthcare systems engineering and have a direct relationship with increasing patient satisfaction and decreasing the healthcare system cost. Various predictive approaches are applied and compared with the proposed multi-task approach in this study. This research contributes developing a more accurate patient workload predictive model to improve prediction effectiveness in different ways. First, to the best of our knowledge, there is no quantitative and statistical model developed for predicting the workload that patients impose to health providers in patient-centered medical home health system while using the RVUs as a quantitative measure for workload prediction. Secondly, a heuristic cluster-based single-task learning algorithm is developed in this research in order to successfully reduce the prediction error. Thirdly, our proposed method is a facility dependent method meaning that similar patients are analyzed while considering the characteristics of their related hospital. Thus, similar patients in different hospital can have different workload due to the difference in hospitals' health systems efficiency, location, etc. Fourthly, in this study relatedness of multiple instances are considered while training the model. So, we tackled the limited training samples issue by integrating related task in the training phase. Last but not least, in this research, a comprehensive multi-task framework is proposed for analyzing the datasets whereby the instances share a hidden information layer but differ in other characteristics.

II. RELATED WORKS

In this section, the previous works on healthcare patient workload prediction by using Relative Value Units (RVUs) and multi-task learning are discussed in detail.

*A. Relative Value Units and Patient Workload Prediction*

Relative value units are considered as a standardized measure of outpatient workload. RVUs are considered as a national standard for measuring productivity, budgeting, allocating resources, and cost benchmarking. RVU represents the relative amount of physician work, resources, and expertise that is necessary for providing healthcare services to patients. for these reasons many researchers use RVUs as a main source to assess patients' demand and workload volume. Therefore, having an accurate workload predictor is necessary to develop a reasonable estimation of future demand and cost for allocating resources appropriately. Most of the conducted RVU prediction studies are limited to descriptive analysis or regression-based modeling for forecasting the workload. Some of the prominent studies that use RVU as a measure of workload are discussed in this section.

In the early stages of studying RVUs, most of the researchers focused on finding the attributes that affect the workload. Moniz [3] uses RVUs to determine the relationship between workload and three different categories of gender, age and beneficiary type, using univariate analysis of variance. The patients are categorized by predisposing characteristics, i.e., age, gender and social structure, then the average RVU for each category is calculated, and the difference between the mean RVU and mean RVU per beneficiary is analyzed by univariate analysis of variance. The results show that age, gender and beneficiary category can provide significant value for predicting the workload.

Many factors such as patient's age, gender, and diagnostic codes can affect the workload. Østbye et al. [4] suggest patient's visit frequency is affected by the type of chronic diseases. Also, Naessens et al. [5] show that the number of chronic conditions in a patient will significantly affect clinical workload and medical cost. There are different variables to measure workload. Relative value units are among the most important KPIs in order to measure workload. RVUs have been used as a predictor factor in some studies. Turrentine et al., [6] studied the attributes of mostly elderly patients who undergo major operations to predict morbidity, mortality, and their risk factors. By using stepwise logistic regression mortality as an independent variable is predicted then the effect of age on the outcome variables is investigated. Also, they identified the risk factors predictive of morbidity and mortality associated with age groups.

As it is mentioned predicting RVUs to have a precise understanding of the patients' demand is critical. Using regression-based methods is a straightforward solution to tackle this problem. Murphy [7] Developed a demand-based forecast for RVU volume using multiple linear regression modeling by using the surveys completed by PCMH team members. The relationship of patient workload and per-encounter independent variables such as age, gender, beneficiary category, provider specialty, evaluation and management code and appointment type is studied in this research while assessing the relationship of each independent variable with workload separately. Shah et al., [8] investigated the correlation of surgical procedures and other measures of surgeon effort and RVU by using linear and multiple logistic regression. They showed that RVUs are correlated with certain measures of surgeon work and patient's attributes, such as the frequency of Serious Adverse Events (SAEs), patient's overall morbidity. Also, they showed RVUs are significant predictors of operative time, length of stay, and SAEs.

There are other methods that researchers used to forecast the RVU. One example can be found in Barnes's [9] study. He used RVUs to forecast the future demand for ten specialty practices. Two models were developed to predict the demand. In the first model, time series model is performed to find the most accurate fitting forecast model. The result shows that the exponential smoothing method has the least mean square error. In the second model, the past usage rate is used to forecast the demand; however, this method only works for short-term prediction since uncontrollable external factors may affect the prediction during the long-time horizon.

RVUs have been used as a variable to represent the healthcare demand. Etzioni et al., [10] predict the impact of aging population on the amount of surgical work. This estimation also has been used to forecast the future workload needed for surgical works by simply multiplying the age-specific surgery incidence rate for each procedure by the corresponding work RVUs. The results show that the aging population leads to significant increase in demand for surgical services. So, a robust methodology to handle the growth in workload while maintaining the quality of care is necessary. Crane et al., [11] proposed a task-based framework entitled "entropy" to evaluate the relationship between workload in the emergency department and the crowding using as relative value units, RVU/h, and patients seen per hour as inputs. Their framework measures some aspects of the workload in emergency departments, such as acuity and efficiency. The workload is considered as an operational complexity which can be defined as the total information obtained from each task during observations in the certain period. Therefore, they estimated the workload of all tasks performed by healthcare providers by using entropy formula and assigned an entropy value to different tasks. Chasan et al., [12] use RVU to assess workload and resources in eye care procedures by using descriptive statistical analysis. Arndt et al. conduct another study to assess the workload [13]. They assess the primary workload using a survey about the perceived workload while considering both face to face and non-face to face encounters. Their results show that regardless of health status, the total workload of panel management activities associated with routine primary care that was not face-to-face was more significant than the total workload associated with face-to-face encounters.

Since RVUs represent the workload, some studies try to optimize the capacity taking the RVUs into account. As an example, Bryce and Christensen [14] consider the patient's demand as a variable workload, therefore by finding the mean and variance of workload during different time frames, they fit normal distributions for demands and try to match the resource capacity efficiently to optimize staffing process. Also, some researchers focus on predicting operations' cost while using RVUs as input in order to develop decision support systems for resource allocation process. Fulton et al., [15] incorporate data envelopment analysis (DEA) efficiency scores into a traditional logarithmic-linear cost function, and they identify RVU which is workload volume and complexity as cost drivers for hospital operations.

### B. Multi-Task Learning

Applications of multi-task learning include many areas such as natural language processing, bioinformatics, computer vision and healthcare informatics. Despite the extensive use of multi-task learning in other areas, there are limited literature of application of multi-task learning in health informatics. In this section we explore the studies on healthcare informatics and bioinformatics that used MTL as an approach for prediction.

MTL is used to predict sequence signals in genes finding by using various types of regularization terms [16]. In another study multitask regression is coupled with co-clustering for associating gene expression data with phenotypic signatures [17]. The results of the study show that multi-task regression outperforms traditional Lasso and Ridge regression models. Liu et al. [18] used multi-task learning to predict the efficiency of cross-platform siRNA after ranking biological features by their joint importance. Mordelet and Vert [19] tried to prioritize disease genes by using multi-task learning and taking advantage of shared information across disease genes. Another example of applying multi-task learning in genetics can be found in [20] where instead of the linear regression model, multi-task learning and multiple output regression models are applied for predicting genetic trait. Also, multi-task Lasso regression with $L_1$ and $L_2$ regularization is applied to discover the correlation between genetic markers in multiple populations. Moreover, MTL is utilized for protein subcellular location prediction to capture the shared information among different organism [21].

MTL is used for developing brain-computer interfaces by sharing a Gaussian prior on parameters of different tasks [22]. In another study, MTL is formulated as multiple kernel learning for MHC-I binding and splice-site prediction. Zhou et al. [23] used multi-task regression problem by considering the prediction at each time point as a task for mini-mental state examination and Alzheimer's disease assessment. They used temporal group Lasso regularization term with two components including an $L_{2,1}$-norm penalty to ensure selection of a small subset of features, and a temporal smoothness term to have a small deviation between two regression models at successive time points. In another study on Alzheimer's disease prediction, a sparse Bayesian multi-task learning algorithm in order to exploit the relationships between neuroimaging measures and cognitive scores is developed [24]. In addition, in some studies, multi-task learning is incorporated with time-series, and the model is used for Alzheimer's disease progression prediction [25].

MTL is applied on other types of problems in healthcare informatics. As in [26], biological images are analyzed with MTL while using deep learning architectures such as convolutional neural networks for data feature representation for improving the model performance. MTL is used to formulate survival analysis as a classification problem where there are multiple related survival prediction tasks [27], [28]. As it is discussed, even though MTL has been extensively studied, there is no existing research incorporating MTL in patient workload prediction models. Studies in healthcare systems engineering while considering the relatedness between patients' instances is limited in this field.

### III. METHODOLOGY

The framework of this study is illustrated in Fig. 1. The approach consists of three phases, namely, data pre-processing, unsupervised learning and supervised learning. The steps are explained in detail in the following.

#### A. Data Pre-processing

The quality of data directly affects the quality of the prediction. So, in the first stage of the research, after extracting the data, data is cleaned by removing outliers and taking care of missing values. Based on the type of missing values, missing values are either removed or imputed by use of the mean of all samples with the same class. Then, categorical variables are transformed into numerical variables by generating dummy variables. Afterward, to reduce data redundancy, the data is scaled by using min-max normalization method.

#### B. Unsupervised learning

Clustering is an essential part of the second phase of the framework, in order to develop the cluster-based approach that is used in this paper. The reasons of utilizing unsupervised learning before predictions include: making the data more structured and generating subsets of the dataset each of which groups similar samples in a smaller dataset as input for prediction model while maximizing the dissimilarity with the rest of the subsets. In this study, K-means clustering is used in order to discover similar samples. The optimum number of clusters is found by using Gap statistic.

#### C. Supervised learning models

The last stage of the framework consists comparing the performance of three different approaches (i.e., applying traditional prediction models, cluster-based approach, and multi-task learning method) in order to select the best model for patient workload prediction. There are parameter tuning and model evaluation steps for each of approaches mentioned above. These two steps run iteratively until their performance no longer increases and the best parameters for each model are attained. Then, the best model will be selected based on the prediction accuracy and performance. In the next sections, the cluster-based prediction approach and multi-task learning approach are explained.

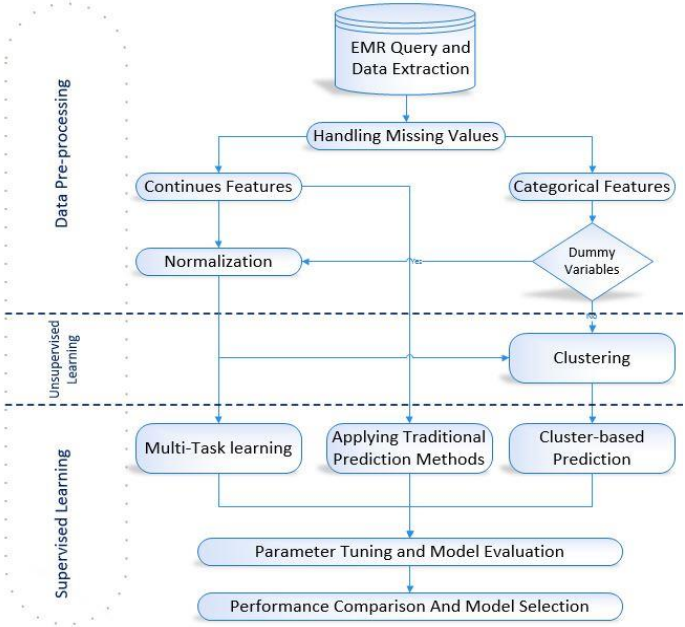

Fig. 1. Patient workload prediction framework

| | **ALGORITHM 1:** CLUSTER-BASED SINGLE-TASK LEARNING |
|---|---|
| 1: | **Input:** a test set of size *N*, set of prediction methods *Q*= {Lasso, Ridge, random forests, regression tree, SVR, XGBoost} |
| 2: | Find the optimum number of clusters (*K*) using the Gap Statistic method |
| 3: | Apply *K*-means clustering |
| 4: | Split data in *K* clusters based on the results of the *K*-means clustering |
| 5: | **for** *i* = 1 to *K* **do** |
| 6: |    **for** *j* = 1 to |*Q*| **do** |
| 7: |       Develop a predictive model for cluster *i* using method *j* |
| 8: |    **end for** |
| 9: | **end for** |
| 10: | **for** *i* = 1 to *N* **do** |
| 11: |    **for** *k* = 1 to *K* **do** |
| 12: |       Compute the distance of record *i* from the center of cluster *k* |
| 13: |    **end for** |
| 14: |    Assign record *i* to the nearest cluster |
| 15: | **end for** |
| 16: | **for** *i* = 1 to *N* **do** |
| 17: |    **for** *j* = 1 to |*Q*| **do** |
| 18: |       Predict the target for record *i* by the model of the associated cluster developed by method *j* |
| 19: |    **end for** |
| 20: | **end for** |
| 21: | **for** *j* = 1 to |*Q*| **do** |
| 22: |    Compute the mean squared error $MSE^j$ for method *j* |
| 23: | **end for** |
| 24: | Pick the results from the method with the minimum $MSE^j$ |

*1) Cluster-based prediction:* The EMR data often contains many patients that are similar in a way. So, ignoring the similarity between patient may reduce the performance of capturing useful patient's information. In order to increase the performance of the first approach (i.e., patient workload prediction by classic machine learning methods), an unsupervised learning approach is added as an essential step before supervised clustering.

In our proposed approach so-called cluster-based prediction, the data is divided into a certain number of clusters by using k-means clustering approach. Eventually, these clusters transform the main dataset into subsets of the primary dataset which include the samples that are closer to each other in terms of the similarity. The objective of the algorithm is to minimize inter-cluster and maximize intra-cluster distance.

As it is illustrated in Fig. 1, one predictive model is adapted specifically for each cluster. That means if the data split into *k* clusters, there would be *k* different predictive models for predicting the workload of a new patient. Therefore, predicting the target value for each patient necessitates developing a mechanism to determine how the new data point belongs to clusters. In order to predict the workload of a new patient, it is necessary to choose one of the developed *k* models. So, Algorithm 1 which is a distance-based algorithm is developed as a mechanism for assigning new patients to their closets cluster. After assigning the patient to the best cluster, the predictive model related to the chosen cluster is used for predicting the patient's workload.

*2) Multi-task learning approach:* In contrary with single task learning, multi-task learning is a paradigm that takes advantage of the relatedness between samples to leverage the knowledge of other related tasks in order to learn a specific task. As it is mentioned in the previous section, studies show that learning multiple tasks jointly instead of individually, results in performance improvement.

Fig. 2 illustrates the difference between multi-task learning and cluster-based single task learning. It shows that single task learning trains each task individually (horizontal learning); however in MTL the tasks are connected, so the hidden layer of information can be shared among the tasks (the vertical relationship between task). Therefore, tasks can affect each other. Furthermore, since in single task learning each task uses its data, when the number of data for each task is not enough, the problem of over-fitting may occur. In this case, MTL can be used to overcome over-fitting by sharing the data over different tasks and increasing the number of samples for each task as it is shown in Fig. 2. MTL can be beneficial since tasks can affect each other and the knowledge between them can be transferred.

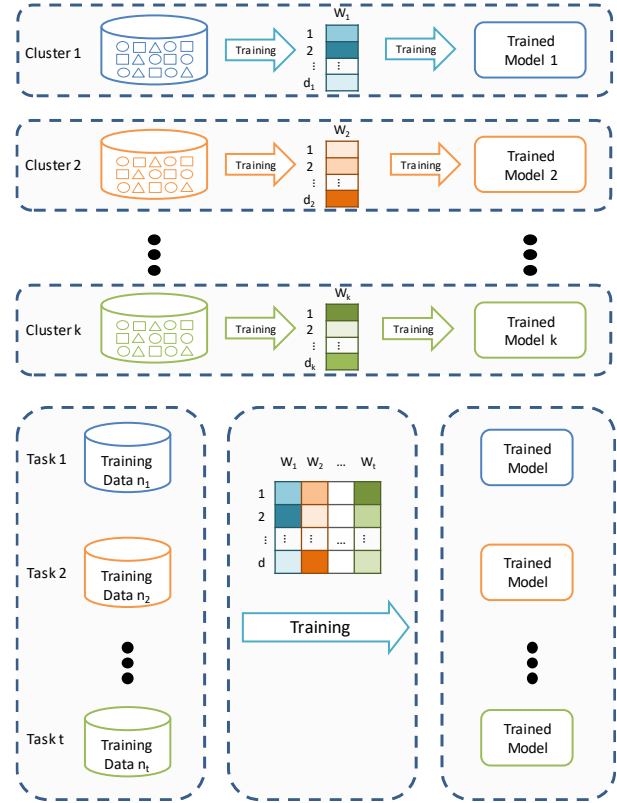

Fig. 2. Comparison beween learning schema of cluster-based single task learning (top) and multi-task learning (bottom)

The objective function of MTL is to minimize the summation of the loss function and task regularization term that is defined as follows.

$$\min_{W} L(W) + \mathcal{R}(W) \quad (1)$$

In (1), $W$ represents the collection of weight vectors that are learned for each task. In this study, the number of tasks and the attributes is represented by $T$ and $D$, respectively. So, the weight matrix is a $T \times D$ matrix ($W \in \mathbb{R}^{T \times D}$). $L(W)$ is the loss function and $\mathcal{R}(W)$ is the regularization term that are expressed in (2) and (3), respectively [29].

$$L(W) = \frac{1}{2} \sum_{t=1}^{T} \|X_t W^T - \beta_t\|_F^2 \quad (2)$$

$$\mathcal{R}(W) = \|W_{2,1}\| = \sum_{d=1}^{D} \sqrt{\sum_{t=1}^{T} |w_{td}|^2} \quad (3)$$

Where $X_t \in \mathbb{R}^{n_t \times D}$ and represents the input data for task $t$. $n_t$ represents the number of samples for each task $t$ and $\beta_t$ is the response value corresponding to the samples in $X_t$. So, the objective function can be written as:

$$\min_{W} = \frac{1}{2} \sum_{t=1}^{T} \|X_t W^T - \beta_t\|_F^2 + \lambda \sum_{d=1}^{D} \sqrt{\sum_{t=1}^{T} |w_{td}|^2} \quad (4)$$

Where $\lambda \geq 0$ is defined as a tuning parameter that bias the data and controls the shrinkage of the model to make the model relativity simpler and sparser to reduce the complexity of the model.

## IV. CASE STUDY AND RESULTS

Many government programs and private payers use the resource-based relative value scale and the relative value unit methodology system as the basis for payment many physician practices. This system is the foundation of medical group financial analysis and is unique to the medical service industry. Relative value units are a national standard used for measuring productivity, budgeting, allocating expenses, and cost benchmarking. RVU represents the relative amount of physician workload.

In this study, patients' data from VA facilities across the nation is used. The dataset contains patient risk factors such as demographic and socioeconomic variables for 5305 samples. The healthcare workload imposed on each provider were measured in relative value unit for one year. The RVU schema has been widely used for reimbursement in VA and Centers for Medicare and Medicaid Services. In this schema, each value was assigned to service (as defined by a coding system called Current Procedural Terminology (CPT) rendered by a provider. The values are adjusted based on geographic regions. One advantage of using RVUs in our approach as opposed to simple face-to-face visit counts lies in its ability to accommodate further workloads that are generated by telephone encounters.

As it is stated in section III, the first part of the proposed approach includes data preprocessing. The pre-processing has been done by performing four steps. First, the missing values are either removed or imputed based on the type of the missing attribute. Then outliers are identified and removed. Afterward, based on the method that is used for prediction categorical variables are converted to a numerical variable by taking advantage of indicator variables. At last, the data is normalized by using min-max normalization method.

After preparing data for analysis, in the supervised learning phase of this research, traditional methods of predictions are adapted in order to predict the RVU for each patient using single task learning. In this research, four well-known machine learning methods for workload prediction are used (i.e., Random Forests, Regression Tree, Lasso and Ridge Regression, Support Vector Regression (SVR), and XGBoost). In order to compare the performance of different methods, Mean Squared Error (MSE) and $R^2$ are considered as the performance metrics for performance evaluation. MSE indicated the squared average deviation of estimated value while $R^2$ represents the goodness of estimator fit. The result of single task learning is presented in Table I.

Cluster-based single task learning is another approach that is proposed and implemented in this study. We take advantage of unsupervised learning by using proposed Algorithm 1 and then single task learning. In the unsupervised part of the cluster-based approach, the gap statistic method is used in order to determine the optimum number of clusters for k-means clustering. As it is indicated in Fig. 3, gap statistic analysis shows that the optimal number of clusters for our dataset is equal to 3.

After determining the optimal number of clusters, and clustering data into 3 clusters four traditional models (i.e., Random Forests, Regression tree, Lasso and Ridge regression, Support Vector Regression (SVR), and XGBoost) are developed for each cluster. Then, Algorithm 1 is used as a heuristic algorithm to assign each element of the test set to a cluster and its corresponding prediction model. So, for each sample in the training set, the distance of the sample from the center of each cluster is computed. Then, the target for the new sample is estimated by the four single task learning models proposed in this research. The performance of the cluster-based approach is demonstrated in Table I.

In the last stage, we implemented multi-task learning. So far, in the above-mentioned approaches the information that is contained between tasks is ignored. However, in multi-task learning, we consider a common hidden layer for two related tasks. That is why multi-task learning has the best performance among other approaches according to the results of table I. We performed multi-task learning on the data for predicting the workload of patients with 10-fold cross-validation for training and testing process. The data is categorized into 130 tasks, which indicate the number of healthcare facilities in the dataset. As in most of EHR data, our dataset contains both patient level and facility level information. So, there is a possibility that many patients have the same attributes but different workloads. The source of this variability can be traced in the facility-level information where the quality and the cost of providing healthcare service, as well as providers diagnosis may differ in different facilities. For example, the cost of a detailed assessment visit with code 99214 performed in San Francisco is higher compared to one performed in Detroit. The other reason that multi-task learning outperforms the rest of approaches is that MTL takes advantage of pooling the sample across the related tasks. So, MTL increases the number of samples for each task in the case that the data is not enough. Thus, not only it avoids overfitting and increases the ability to fit random noise by introducing inductive bias (regulating the model), but also it helps the model to achieve better representation and generalization compare to single-task learning approaches as the results show in Table I.

As the results of Table I suggest, by increasing the number of samples for each task the performance of the prediction models gets closer to each other. Thus, the gap between the performance of the multi-task learning approach and the other approaches is not significant. This complies with one of the main characteristics of the multi-task learning which is the ability of this method in handling the issue of the limited number of samples in the dataset for prediction. Therefore, as the results suggest the main competitive advantage of the multi-task leaning is pooling the sample across the related tasks so that it is able to use the similar samples effectively to increase the sample size of each task. However, when the number of samples for each task is relatively enough for the prediction model to make the accurate prediction, the mentioned benefit is not as significant as when the number of samples are limited.

TABLE I. PERFORMANCE COMPARISON OF THE PROPOSED FRAMEWORK ON THE PCMH DATASET

| Data Type | Method | Measure | Single Task Learning | | Cluster-based Single Task Learning | | Multi-task Learning | |
|---|---|---|---|---|---|---|---|---|
| | | | *MSE* | *$R^2$* | *MSE* | *$R^2$* | *MSE* | *$R^2$* |
| PCMH Dataset | Lasso Regression | Mean | 8.12 | 0.3 | 7.35 | 0.37 | 5.2 | 0.55 |
| | | Var. | 0.26 | 0.0082 | 0.21 | 0.0078 | 0.42 | 0.0029 |
| | Ridge Regression | Mean | 6.61 | 0.43 | 5.88 | 0.48 | | |
| | | Var. | 0.22 | 0.0071 | 0.28 | 0.0075 | | |
| | Regression Tree | Mean | 9.87 | 0.2 | 7.71 | 0.34 | | |
| | | Var. | 0.31 | 0.0069 | 0.46 | 0.0054 | | |
| | Random Forests | Mean | 7.64 | 0.35 | 6.93 | 0.4 | | |
| | | Var. | 0.29 | 0.0093 | 0.33 | 0.0069 | | |
| | SVR | Mean | 10.82 | 0.15 | 7.97 | 0.32 | | |
| | | Var. | 0.31 | 0.013 | 0.39 | 0.0085 | | |
| | XGBoost | Mean | 7.4 | 0.36 | 6.82 | 0.41 | | |
| | | Var. | 0.25 | 0.0055 | 0.25 | 0.0063 | | |
| Simulated Dataset | Lasso Regression | Mean | 6.24 | 0.21 | 5.01 | 0.35 | 4.02 | 0.48 |
| | | Var. | 0.31 | 0.0092 | 0.34 | 0.0026 | 0.31 | 0.0065 |
| | Ridge Regression | Mean | 5.37 | 0.32 | 4.44 | 0.43 | | |
| | | Var. | 0.39 | 0.0053 | 0.52 | 0.0067 | | |
| | Regression Tree | Mean | 5.22 | 0.34 | 4.78 | 0.39 | | |
| | | Var. | 0.42 | 0.0071 | 0.26 | 0.0053 | | |
| | Random Forests | Mean | 4.83 | 0.41 | 4.2 | 0.46 | | |
| | | Var. | 0.33 | 0.0089 | 0.38 | 0.0084 | | |
| | SVR | Mean | 6.28 | 0.2 | 5.21 | 0.33 | | |
| | | Var. | 0.27 | 0.0048 | 0.35 | 0.0079 | | |
| | XGBoost | Mean | 4.27 | 0.45 | 3.92 | 0.49 | | |
| | | Var. | 0.22 | 0.0064 | 0.29 | 0.0033 | | |

In order to validate the results of this study, the proposed approach is implemented on a simulated dataset. We benefit from data simulation to show the efficiency of the proposed method where the features follow the same distribution as the main dataset but with larger number of samples in each task. Considering the distribution of each feature we generated a larger dataset with 130,000 data samples. We generally followed the proposed approach on the simulated data. The single-task learning, clustered-based single task learning and multi-task learning approaches are applied to the simulated data. In order to obtain robust results, in the first step parameters of each method are fine-tuned. Then, we replicate each algorithm 50 times. In the next step, the average and variance of the MSE and the corresponding $R^2$ are calculated and shown in Table I.

V. CONCLUSION AND DISCUSSION

In this study, a framework for patient demand for service prediction in a multi-facility environment, where the data instances are inter-related, and the dataset contains both patient-dependent and facility-dependent attributes. The main limitation of previous studies in healthcare workload prediction is that the problem is not investigated for multiple facilities where there is a difference between the quality of provided service, the equipment, and resources used for provided service as well as the diagnosis and treatment procedures in each facility for patients with similar conditions.

The proposed approach in this research investigates the effect of incorporating the relatedness between the tasks into the model on the performance of the predictive model. In order to attain this goal, after data preprocessing, three approaches implemented on data. In the first step, standard supervised single-task learning methods are adapted on the data for demand prediction whit no consideration of relatedness. In the second step, before supervised learning, we have applied k-means clustering on the data (unsupervised learning) for clustering the patients with similar attributes into a certain number of groups. Then, by using algorithm 1, the new test instances are assigned to the closet clusters, and their target value is computed based on the assigned predictive model of each cluster. So, by developing the approach, we considered the features relatedness to some extent. In the end, we used MTL to model the problem and consider the task relatedness and information transferring between the tasks. The results show that considering the task relatedness leads to improve the performance of the predictive model.

In order to justify the feasibility of the approach, we compared the results of classic machine learning techniques with our proposed approach. Due to the relatedness between the tasks, by developing the algorithm mentioned above, we aim to make the similar instances as close as possible and group them into a particular number of clusters. Then a predictive model that represents each cluster is adapted to predict the target value for new patients by assigning them to the predictive model of the most similar cluster. For the experimental study, we used the patients' data from 130 VA facilities across the US. The dataset contains the patient demographic and disease information as well as facility information as input features and the relative workload as the target. In addition, a heuristic algorithm for cluster-based single task learning is developed in this study. After implementing different approaches, the performance of each method is compared, and the model with the best performance is selected which is MTL in our case study.

There are many reasons that MTL outperformed other approaches in this study. MTL uses auxiliary tasks to learn model as a form of inductive transfer. In addition, MTL implicitly augments data of related task that results in an increase in number of instances for each task that avoids overfitting. Also, since MTL learns multiple tasks simultaneously and incorporates different noise patterns of different tasks, it can learn a more general representation. Furthermore, when the data is limited, noisy and high-dimensional, MTL differentiate between relevant and irrelevant attributes by using the additional information that are provided by other tasks. Moreover, in order to improve the generalization potential of the model, while MTL learns the model, it biases the model to a representation that is preferable for other tasks. Also, as it is mentioned, MTL increases the ability to fit random noise by regulating the model.

In summary, as it is discussed imposing inductive bias to the model to make the model able to transfer the information between related tasks can improve the performance of the predictive model when there are limited instances for some tasks in the dataset. To the best of our knowledge, this study is the first study that tried to incorporate the relatedness of tasks into the patient demand prediction models by developing cluster-based single task learning method and implementing MTL. The results of this study show that considering the shared representation of hidden layer of data for inter-related tasks and training all the task jointly increase the performance of the prediction. As a future direction, we suggest the implementation of feature selection, representation and learning methods on the data to summarize and transform the data into better shape and capture more information from the abstracted data before implementing predictive models.

ACKNOWLEDGMENT

This research is a part of a research supported by the National Science Foundation, Division of Civil, Mechanical, and Manufacturing Innovation (CMMI) under grant number 1233504.

REFERENCES

[1] C. Aguwa, M. H. Olya, and L. Monplaisir, "Modeling of fuzzy-based voice of customer for business decision analytics," *Knowledge-Based Syst.*, vol. 125, pp. 136–145, Jun. 2017.

[2] Y. Zhang and Q. Yang, "A survey on multi-task learning," *arXiv Prepr. arXiv1707.08114*, 2017.

[3] C. Moniz, "Outpatient workload (RVU) predictors: age, gender & beneficiary category," JOHNS HOPKINS UNIV BALTIMORE MD, 2008.

[4] T. Østbye, K. S. H. Yarnall, K. M. Krause, K. I. Pollak, M. Gradison, and J. L. Michener, "Is there time for management of patients with chronic diseases in primary care?," *Ann. Fam. Med.*, vol. 3, no. 3, pp. 209–214, 2005.

[5] J. M. Naessens, R. J. Stroebel, D. M. Finnie, N. D. Shah, A. E. Wagie, W. J. Litchy, P. J. Killinger, T. J. O'Byrne, D. L. Wood, and R. E. Nesse, "Effect of multiple chronic conditions among working-age adults," *Am. J. Manag. Care*, vol. 17, no. 2, pp. 118–122, 2011.

[6] F. E. Turrentine, H. Wang, V. B. Simpson, and R. S. Jones, "Surgical risk factors, morbidity, and mortality in elderly patients," *J. Am. Coll. Surg.*, vol. 203, no. 6, pp. 865–877, 2006.

[7] R. G. Murphy, "A Primary Care Workload Production Model for Estimating Relative Value Unit Output," DTIC Document, 2011.

[8] D. R. Shah, R. J. Bold, A. D. Yang, V. P. Khatri, S. R. Martinez, and R. J. Canter, "Relative value units poorly correlate with measures of surgical effort and complexity," *J. Surg. Res.*, vol. 190, no. 2, pp. 465–470, 2014.

[9] T. D. Barnes, "Demand Analysis for Proposed Medical Services at the Future Naval Health Clinic Charleston, South Carolina: A Graduate Management Project," DTIC Document, 2006.

[10] D. A. Etzioni, J. H. Liu, M. A. Maggard, and C. Y. Ko, "The aging population and its impact on the surgery workforce," *Ann. Surg.*, vol. 238, no. 2, pp. 170–177, 2003.

[11] P. W. Crane, Y. Zhou, Y. Sun, L. Lin, and S. M. Schneider, "Entropy: A conceptual approach to measuring situation-level workload within emergency care and its relationship to emergency department crowding," *J. Emerg. Med.*, vol. 46, no. 4, pp. 551–559, 2014.

[12] J. E. Chasan, B. Delaune, A. Y. Maa, and M. G. Lynch, "Effect of a teleretinal screening program on eye care use and resources," *JAMA*

*Ophthalmol.*, vol. 132, no. 9, pp. 1045–1051, 2014.

[13] B. Arndt, W.-J. Tuan, J. White, and J. Schumacher, "Panel Workload Assessment in US Primary Care: Accounting for Non–Face-to-Face Panel Management Activities," *J. Am. Board Fam. Med.*, vol. 27, no. 4, pp. 530–537, 2014.

[14] D. J. Bryce and T. J. Christensen, "Finding the sweet spot: how to get the right staffing for variable workloads: a simulation tool can help hospitals uncover hidden opportunities to reduce costs by optimizing staffing in a way that best reflects demand," *Healthc. Financ. Manag.*, vol. 65, no. 3, pp. 54–61, 2011.

[15] L. Fulton, L. S. Lasdon, and R. R. McDaniel, "Cost drivers and resource allocation in military health care systems," *Mil. Med.*, vol. 172, no. 3, pp. 244–249, 2007.

[16] C. Widmer, J. Leiva, Y. Altun, and G. Rätsch, "Leveraging sequence classification by taxonomy-based multitask learning," in *Annual International Conference on Research in Computational Molecular Biology*, 2010, pp. 522–534.

[17] K. Zhang, J. W. Gray, and B. Parvin, "Sparse multitask regression for identifying common mechanism of response to therapeutic targets," *Bioinformatics*, vol. 26, no. 12, pp. i97–i105, 2010.

[18] Q. Liu, Q. Xu, V. W. Zheng, H. Xue, Z. Cao, and Q. Yang, "Multi-task learning for cross-platform siRNA efficacy prediction: an in-silico study," *BMC Bioinformatics*, vol. 11, no. 1, p. 181, 2010.

[19] F. Mordelet and J.-P. Vert, "ProDiGe: Prioritization Of Disease Genes with multitask machine learning from positive and unlabeled examples," *BMC Bioinformatics*, vol. 12, no. 1, p. 389, 2011.

[20] D. He, D. Kuhn, and L. Parida, "Novel applications of multitask learning and multiple output regression to multiple genetic trait prediction," *Bioinformatics*, vol. 32, no. 12, pp. i37–i43, 2016.

[21] Q. Xu, S. J. Pan, H. H. Xue, and Q. Yang, "Multitask learning for protein subcellular location prediction," *IEEE/ACM Trans. Comput. Biol. Bioinforma.*, vol. 8, no. 3, pp. 748–759, 2011.

[22] M. Alamgir, M. Grosse–Wentrup, and Y. Altun, "Multitask learning for brain-computer interfaces," in *Proceedings of the Thirteenth International Conference on Artificial Intelligence and Statistics*, 2010, pp. 17–24.

[23] J. Zhou, L. Yuan, J. Liu, and J. Ye, "A multi-task learning formulation for predicting disease progression," in *Proceedings of the 17th ACM SIGKDD international conference on Knowledge discovery and data mining*, 2011, pp. 814–822.

[24] J. Wan, Z. Zhang, J. Yan, T. Li, B. D. Rao, S. Fang, S. Kim, S. L. Risacher, A. J. Saykin, and L. Shen, "Sparse Bayesian multi-task learning for predicting cognitive outcomes from neuroimaging measures in Alzheimer's disease," in *Computer Vision and Pattern Recognition (CVPR), 2012 IEEE Conference on*, 2012, pp. 940–947.

[25] H. Wang, F. Nie, H. Huang, J. Yan, S. Kim, S. Risacher, A. Saykin, and L. Shen, "High-order multi-task feature learning to identify longitudinal phenotypic markers for alzheimer's disease progression prediction," in *Advances in Neural Information Processing Systems*, 2012, pp. 1277–1285.

[26] W. Zhang, R. Li, T. Zeng, Q. Sun, S. Kumar, J. Ye, and S. Ji, "Deep model based transfer and multi-task learning for biological image analysis," *IEEE Trans. Big Data*, 2016.

[27] L. Wang, Y. Li, J. Zhou, D. Zhu, and J. Ye, "Multi-task Survival Analysis," in *2017 IEEE International Conference on Data Mining (ICDM)*, 2017, pp. 485–494.

[28] Y. Li, J. Wang, J. Ye, and C. K. Reddy, "A multi-task learning formulation for survival analysis," in *Proceedings of the 22nd ACM SIGKDD International Conference on Knowledge Discovery and Data Mining*, 2016, pp. 1715–1724.

[29] R. Tibshirani, "Regression shrinkage and selection via the lasso," *J. R. Stat. Soc. Ser. B*, pp. 267–288, 1996.